\documentclass[12pt]{article}

\def\rf#1{(\ref{eq:#1})}
\def\lab#1{\label{eq:#1}}

\def\br{\begin{eqnarray}}
\def\er{\end{eqnarray}}
\def\be{\begin{equation}}
\def\ee{\end{equation}}
\def\({\left(}
\def\){\right)}

\def\rlx{\relax\leavevmode}
\def\IR{\rlx\hbox{\rm I\kern-.18em R}}
\def\vp{\varphi}
\def\u2{\mid u\mid^2}

\def\IZ{\rlx\hbox{\sf Z\kern-.4em Z}}
\def\IR{\rlx\hbox{\rm I\kern-.18em R}}
\def\IC{\rlx\hbox{\,$\inbar\kern-.3em{\rm C}$}}

\input{epsf}

\usepackage{graphics}

\begin{document}

\begin{titlepage}
\vspace*{-1cm}

\vskip 3cm

\vspace{.2in}
\begin{center}
{\large\bf Exact vortex solutions in an extended Skyrme-Faddeev model}
\end{center}

\vspace{.5cm}

\begin{center}
L. A. Ferreira\footnote{e-mail: {\tt laf@ifsc.usp.br}}

\vspace{.5 in}
\small

\par \vskip .2in \noindent
Instituto de F\'\i sica de S\~ao Carlos; IFSC/USP;\\
Universidade de S\~ao Paulo - USP \\ 
Caixa Postal 369, CEP 13560-970, S\~ao Carlos-SP, Brazil\\

\normalsize
\end{center}

\vspace{.5in}

\begin{abstract}
We construct exact vortex solutions in $3+1$ dimensions to a theory
which is an extension, 
due to Gies, of the Skyrme-Faddeev model, and that is believed to
describe some aspects of the low energy limit of the pure $SU(2)$
Yang-Mills theory. Despite the efforts in the last decades those are
the first exact analytical solutions to be constructed for such type
of theory. The exact vortices appear in a very par\-ti\-cu\-lar sector
of the theory characterized by  special values of the 
coupling constants,
and by a constraint that leads to an infinite number of conserved 
charges.  The theory is
scale invariant in that sector, and  the solutions satisfy Bogomolny type
equations. The energy of 
the static
vortex is proportional to its topological charge, and waves can
travel with the speed of light along them, adding to the energy
a term  proportional to a $U(1)$ Noether charge they create. 
We believe such
vortices may play a role in the strong coupling regime of the pure
$SU(2)$ Yang-Mills theory. 
\end{abstract}

%\pacs{11.27.+d, 11.10.Lm, 11.15.-q,  11.30.-j}

\end{titlepage}

The Skyrme-Faddeev  (SF) model \cite{sf} has been proposed long ago as
a four dimensional field theory that can present finite energy knot solitons
carrying a topological charge given by the Hopf map $S^3\rightarrow
S^2$. Despite many efforts in the last decades no analytical 
non-trivial solutions have been constructed. The confirmation of that
expectation however came some years ago with the construction of numerical
solutions employing powerful computer resources
\cite{sutcliffe,hietarinta,solfn}. The interest in the SF model has 
increased since then and many applications have been proposed including
Bose-Einstein condensates \cite{babaev1} and superconductors
\cite{babaev2}. In addition, it
has been conjectured \cite{fn} that the SF model describes the low
energy limit of the pure (no matter) $SU(2)$ Yang-Mills theory, with
the knot solitons being interpreted perhaps as glueballs, or some
non-trivial vacuum configurations. The conjecture is based on 
the Cho-Faddeev-Niemi decomposition \cite{chofn,fn} of the $SU(2)$
Yang-Mills field 
in terms of an abelian gauge field,  a
triplet ${\vec n}$ of scalars living on $S^2$ (${\vec n}^2=1$), and
two other degrees of freedom which can be expressed by two real scalar
fields. There 
has been several controversies about the validity of that conjecture
\cite{wipf,newfaddeev} but a
calculation carried out by Gies \cite{gies} has indicated that it may
indeed be the case provided additional quartic terms are added to the
effective Lagrangian.  

The purpose of  this paper is to present the first exact analytical
solutions in $3+1$ dimensions for the extension of the Skyrme-Faddeev
model defined by the Lagrangian   
\be
{\cal L} = M^2\, \partial_{\mu} {\vec n}\cdot\partial^{\mu} {\vec n}
 -\frac{1}{e^2} \, \(\partial_{\mu}{\vec n} \wedge 
\partial_{\nu}{\vec n}\)^2 + \frac{\beta}{2}\,
\left(\partial_{\mu} {\vec n}\cdot\partial^{\mu} {\vec n}\right)^2
\lab{action}
\ee
where ${\vec n}$ is a triplet of real scalar fields taking values on
the sphere $S^2$, $M$ is a coupling constant with dimension of mass,
 $e^2$ and $\beta$ are dimensionless coupling constants. We show that
\rf{action} possesses a Bogomolny type sector admitting an infinite number of
conserved currents, and when the coupling constants satisfy $\beta\,
e^2=1$, there exists an infinite class of exact solutions. Among those
there are static vortex solutions as well as vortices with waves
travelling along them, and enhancing their stability.   

The first two
 terms of \rf{action} correspond to the  original SF  
 model \cite{sf,solfn}, and the third one  agrees with  the term found by
 Gies in his calculations \cite{gies}. A similar extension of the SF 
 model  has been considered in \cite{glad}, but with some important
 differences. First the model was defined in three space dimensions
 only and the signs of the second and third terms in \rf{action} were
   chosen to keep the static Hamiltonian positive definite. The
   existence of the exact solutions we present in this paper depends
   crucially on the fact  that $e^2$ and $\beta$ have the same
   sign, and that agrees with the signature of the one-loop Wilsonian
   effective action for the $SU(2)$ Yang-Mills theory obtained in
   \cite{gies}. Despite of that signature, our solutions  
   belong to a sector of the theory where the energy is positive
   de\-fi\-ni\-te.  Solutions for the conformally
invariant version of the model, namely the theory \rf{action} without
the first term ($M=0$), have been constructed and will be given in a
separate paper \cite{andre}. 

The solutions we construct in this paper can be best explained by
using the stereographic projection of $S^2$ in terms of a complex
scalar field $u$, i.e. 
\be
{\vec n} = \(u+u^*,-i\(u-u^*\),\u2 -1\)/\(1+\u2\)
\ee
One then obtains that 
\br
{\vec n}\cdot\(\partial_{\mu}{\vec n} \wedge 
\partial_{\nu}{\vec n}\)
&=&2i\,\frac{\(\partial_{\nu} u \partial_{\mu} u^* - 
\partial_{\mu} u\partial_{\nu} u^*\)}{\(1+\u2\)^2} 
\nonumber\\
\left(\partial_{\mu} {\vec
  n}\cdot \partial^{\mu} {\vec
  n}\right)&=&4\,\frac{\partial_{\mu}u\;\partial^{\mu}u^*}{\(1+\u2\)^2}
\lab{dn2}
\er
Therefore, the Lagrangian \rf{action} becomes
\be
{\cal L}=
4\,M^2\,\frac{\partial_{\mu}u\;\partial^{\mu}u^*}{\(1+\u2\)^2} + 
\frac{8}{e^2}\left[ 
\frac{\(\partial_{\mu}u\)^2\(\partial_{\nu}u^*\)^2}{\(1+\u2\)^4}+
\(\beta\,e^2-1\)\,\frac{\(\partial_{\mu}u\;\partial^{\mu}u^*\)^2}{\(1+\u2\)^4}
\right]
\lab{actionu}
\ee
and the Euler-Lagrange equations following from \rf{actionu}, or
\rf{action},  reads  
\be
\(1+\u2\)\, \partial^{\mu}{\cal K}_{\mu}-2\,u^{*}\,{\cal K}_{\mu}\,
\partial^{\mu} u=0
\lab{eqmot}
\ee
together with its complex conjugate, and where\footnote{Notice that
  this theory has a set of trivial plane wave solutions given by $u=A
  e^{i\,k_{\mu}\,x^{\mu}}$, with the wave vector satisfying either
  $k^2=0$ or
  $k^2=-\frac{\(1+A^2\)^2}{4\,A^2}\,\frac{M^2}{\beta}$. Plane wave
  solutions for the SF model were considered in \cite{hirayama}.}
\be
{\cal K}_{\mu}\equiv M^2\, \partial_{\mu}u 
-\frac{4}{e^2}\,\frac{ 
\left[\(1-\beta\,e^2\)\,\(\partial_{\nu}u\,\partial^{\nu}u^{*}\)\,
\partial_{\mu} u-
\(\partial_{\nu}u\partial^{\nu} u\)
\partial_{\mu}u^{*}\right]}{\(1+\u2\)^2}
\lab{kdef}
\ee

It was shown in \cite{afs}, using a generalization of the zero
curvature condition to higher dimensions, that many field theories
admit an infinite number of conserved charges when some special
constraints are imposed. In theories with target space $S^2$ like
\rf{action} the relevant constraint is given by
\be
\partial_{\mu}u\, \partial^{\mu}u=0
\lab{constraint}
\ee
When \rf{constraint} is imposed the theory \rf{action} possesses the
infinite set of  conserved currents  given by 
\be
J_{\mu}\equiv \frac{\delta G}{\delta u}\, {\cal K}^{c}_{\mu}   - 
\frac{\delta G}{\delta u^{*}}\, {{\cal K}^{c}_{\mu}}^{*}
\lab{currents}
\ee
where $G$ is any function of $u$ and $u^{*}$, but not of
their derivatives, and ${\cal K}^{c}_{\mu}$ is obtained from \rf{kdef}
by imposing \rf{constraint}, i.e.
\be
{\cal K}^{c}_{\mu}\equiv M^2\, \partial_{\mu}u-\frac{4}{e^2}\,\frac{ 
\(1-\beta\,e^2\)\,\(\partial_{\nu}u\,\partial^{\nu}u^{*}\)\,
\partial_{\mu} u}{\(1+\u2\)^2}
\lab{kcdef}
\ee
 The conservation of \rf{currents} follows
from the equations of motion, which now read $\partial^{\mu}{\cal
  K}^{c}_{\mu}=0$, and the two identities ${\cal
  K}^c_{\mu}\partial^{\mu}u=0$ and $Im\({\cal
  K}^{c}_{\mu}\partial^{\mu}u^*\)=0$. In the context of SF model those
currents were first constructed in \cite{afzplb}.  

Notice that the constraint \rf{constraint}, in Minkowiski space-time,
can be cast into the form  
\be
\left[\(\partial_1+i\,\partial_2\) u\right]
\left[\(\partial_1-i\,\partial_2\) u\right] = -
\left[\(\partial_3+\partial_0\)u\right] 
\left[\(\partial_3-\partial_0\)u\right] 
\lab{bogo1}
\ee
If one now chooses the coupling constants in \rf{action} to satisfy
\be
\beta\,e^2=1
\lab{bogo3}
\ee
one gets that eqs. of motion \rf{eqmot} reduce to
$\partial^2u=0$, or 
\be
 \(\partial_1+i\,\partial_2\)
\(\partial_1-i\,\partial_2\) u = - \(\partial_3+\partial_0\)
\(\partial_3-\partial_0\)u 
\lab{bogo2} 
\ee
In a Euclidean space-time the same relations hold by replacing the
time coordinate $x^0$ by an imaginary time $-i\,x^4$, and so
$\partial_0\rightarrow i\,\partial_4$. 

Of course the equations \rf{bogo1} and \rf{bogo2} are solved by field
configurations satisfying
\be
\(\partial_1+i\,\varepsilon_1\,\partial_2\)u=0 \qquad {\rm and}
\qquad 
\(\partial_3+\varepsilon_2\,\partial_0\)u=0
\lab{bogo4}
\ee
where $\varepsilon_i=\pm 1$, are signs chosen independently. In
addition, equations \rf{bogo4} are satisfied by configurations of the
form\footnote{We are grateful to Prof. D. Fairlie for pointing to us that a
  large class of implicit solutions for the wave 
  equation and the constraint \rf{constraint} is given in
  \cite{fairlie}, and that they are related to the solutions \rf{ansatzu}.}
\be
u=v\(z\)\, w\(y\)
\lab{ansatzu}
\ee
where $z=x^1+i\,\varepsilon_1\,x^2$ and $y=x^3-\varepsilon_2\,x^0$,
with $v\(z\)$ and $w\(y\)$ being arbitrary regular functions of their
arguments. Notice that if $u$ satisfies \rf{bogo4}, so does any
regular functional of it, ${\cal F}\(u\)$. Indeed, by taking ${\cal
  F}$ to be the logarithm one observes that the ansatz \rf{ansatzu} is 
mapped into $u=v\(z\) + w\(y\)$. See \cite{ward} for similar
discussions in $2+1$ dimensions.
 
Notice that by imposing the constraint \rf{constraint} and the
condition \rf{bogo3} the Lagrangian \rf{actionu} and equations of
motion \rf{eqmot} becomes those of the $CP^1$ model, i.e.
\be
{\cal L}_{CP^1}=
4\,M^2\,\frac{\partial_{\mu}u\;\partial^{\mu}u^*}{\(1+\u2\)^2} 
\qquad\qquad \qquad\qquad
\(1+\u2\)\, \partial^2 u-2\,u^{*}\,\(\partial^{\mu} u\)^2=0
\lab{cp1model}
\ee
Therefore, the solutions of \rf{bogo1} and \rf{bogo2} also solve
the $CP^1$ model in $3+1$ 
dimensions. Therefore, the Skyrme-Faddeev-Gies model \rf{action}  
shares a common submodel with $CP^1$, defined by the equations
$\(\partial_{\mu}u\)^2=0$, and $\partial^2u=0$, together with the
condition \rf{bogo3}. In fact, the first equation in
\rf{bogo4} correspond to the Bogomolny equation, or equivalently the
Cauchy-Riemann equations, leading to static solutions of that model in
$2+1$ dimensions \cite{bp}.  The relation between the Bogomolny
equation and the constraint \rf{constraint} was already pointed out in
\cite{afs} when the generalization of the zero curvature condition was
applied to the $CP^1$ model in $2+1$ dimensions. 

The structure of the solutions is now clear. The function $v\(z\)$
corresponds to the static lumps of the $CP^1$ model in $2+1$
dimensions. The function $w\(y\)$ corresponds to waves traveling
along the $x^3$ direction with the speed of light. In a Euclidean
space-time $w\(y\)$ corresponds instead to a second copy of the $CP^1$
lumps. If one wants static solutions in Minkowiski space-time, then it
has also to be $x^3$ independent. Therefore, the $CP^1$ lump gets a
string like shape in three space dimensions and it corresponds in fact
to a static vortex solution for the Skyrme-Faddeev-Gies model
(alternatively for the $CP^1$ too, in three space dimensions).  One
can take $v\(z\)$ to be any finite energy lump solution of the two
dimensional $CP^1$ model. For
instance, one can take $v\sim z^n$, and $w\(y\)=1$ in \rf{ansatzu}. Using polar
coordinates on the $x^1\, x^2$ plane, i.e.  
$x^1 + i\, \varepsilon_1\, x^2=\rho\, e^{ i\,\varepsilon_1\, \vp}$, we
obtain the static vortex 
\be
u= \(\frac{\rho}{a}\)^{n}\; e^{i\, \varepsilon_1\, n\, \vp}
\lab{staticvortex}
\ee
where $n$ is an integer, and $a$ an arbitrary parameter with dimension
of length. Configurations with several vortices all parallel to the
$x^3$-axis can be obtained from the multi lumps of the $CP^1$ model.
Numerical and approximate vortex solutions for the SF model have been
considered in \cite{hirayama}.  

One can dress such vortices with waves traveling along the $x^3$
axis. There are several ways of doing it. One way of keeping the energy
per unit of length finite is to take $w\(y\)$ in \rf{ansatzu} of the
plane wave form, leading to the vortex 
\be
u= \(\frac{\rho}{a}\)^{n}\; e^{i\left[ \varepsilon_1\, n\, \vp+
  k\,\(x^3-\varepsilon_2 \, x^0\)\right]}
\lab{vortexwave}
\ee
where $k$ is an arbitrary parameter with dimension of
(length)$^{-1}$. Vortices with waves have been found in a different
context in \cite{nitta}. 

If one evaluates the Hamiltonian density for the theory \rf{action}
and then imposes the conditions \rf{bogo1} and \rf{bogo3} one obtains 
\be
{\cal H}_c = 4\, M^2\, \frac{\(\partial_0\,u\, \partial_0\,
  u^*+ {\vec \nabla} u\cdot {\vec \nabla} u^*\)}{\(1+\mid u \mid^2\)^2}
\lab{hcdef}
\ee
where the subindex $c$ means we are imposing \rf{bogo1} and
\rf{bogo3}. Notice that it coincides with the Hamiltonian density for
the $CP^1$ model. We can write that as
\br
&{\cal H}_c& = \frac{4 M^2}{\(1+\mid u \mid^2\)^2}\left[ 
\mid \partial_1 u +i\varepsilon_1\partial_2 u\mid^2 +
\mid \partial_3 u +\varepsilon_2\partial_0 u\mid^2 \right. \nonumber\\
&+& \left. 
i\varepsilon_1\(\partial_1 u\partial_2 u^*-\partial_2
u\partial_1 u^*\)
- \varepsilon_2\(\partial_3 u\partial_0 u^*+\partial_0
u\partial_3 u^*\)\right]\nonumber\\
&\geq &
4\, M^2\,\left[ i\,\varepsilon_1 \frac{\(\partial_1 u\,\partial_2
    u^*-\partial_2 
u\,\partial_1 u^*\)}{\(1+\mid u \mid^2\)^2}
\right.\nonumber\\
&-& \left. \varepsilon_2\frac{\(\partial_3 u\,\partial_0 u^*+\partial_0
u\,\partial_3 u^*\)}{\(1+\mid u \mid^2\)^2}\right]
\lab{hcbound}
\er
The bound on the energy density is clearly saturated by the solutions
of \rf{bogo4}.  In fact, for those solutions one gets 
\be
{\cal H}_c = 8\,M^2\,\(\mid \partial_z u\mid^2 +  \mid \partial_y
  u\mid^2\)/\(1+\mid u \mid^2\)^2
\ee
The same analysis can be done in the Euclidean case, with the same
results, replacing the Hamiltonian density by the Lagrangian density, and
$x^0\rightarrow -i\,x^4$.

The energy density for the vortex solutions \rf{staticvortex} and
\rf{vortexwave} is then given by\footnote{Notice that such energy
  density depends on $\mid n\mid$, and not on the sign of $n$, because 
$\frac{\(\rho/a\)^{2\,n}}{\(1+\(\rho/a\)^{2\,n}\)^2}=
\frac{\(\rho/a\)^{-2\,n}}{\(1+\(\rho/a\)^{-2\,n}\)^2}$}
\be
{\cal H}_c^{{\rm vortex}} =
8\,M^2\,\left[\frac{n^2}{\rho^2}+k^2\right]\,
\frac{\(\rho/a\)^{2\,n}}{\(1+\(\rho/a\)^{2\,n}\)^2}  
\ee
Therefore, by integrating on the $x^1x^2$ plane we get the energy per
unity lenght. For the static vortex \rf{staticvortex} it is 
\be
{\cal E}_{\rm stat. vortex} = 8\,\pi\,M^2\,\mid n\mid
\ee
The integer $n$ is the topological charge associated to the vortex,
and defined as the winding number of the map from any circle on the
$x^1x^2$ plane, centered
at the $x^3$-axis, to the circle $u/\mid u\mid$ on target
space. So, the energy per unit length is proportional to the
topological charge, a characteristic of solutions saturating the
Bogomolny bound. 

The energy per unit length for the time-dependent vortex
\rf{vortexwave} diverges for 
$n=\pm 1$, and for $\mid n\mid >1$ it is given  by
\be
{\cal E}_{\rm vortex/wave} = 8\,\pi\,M^2\,\left[ \mid
  n\mid+k^2\,a^2\,I\(\mid n\mid\)\right]
\lab{energyvortexwave}
\ee
where 
$I\(n\) = \frac{1}{n}\,
\Gamma\(\frac{n+1}{n}\)\,\Gamma\(\frac{n-1}{n}\)$, with $\Gamma$ being
the Euler's Gamma function.  
Notice that $I\(n\)$ is a monotonically decreasing function of $n$ and
$I\(n\)\rightarrow 1/n$, as $n\rightarrow \infty$. Therefore, the
plane wave contribution to the energy per unit length decreases as 
$\mid n\mid$ increases. 

Notice that a necessary condition for the charge densities $J_0$
associated to \rf{currents} to be  non-vanishing is that the field $u$
should be time dependent. Therefore, all 
the charges associated to \rf{currents} vanish when evaluated on the
static vortex solutions \rf{staticvortex}. However, the vortices
\rf{vortexwave}, with waves travelling along them, are time dependent
solutions and so can have non-vanishing charges associated to
\rf{currents}. Consequently that infinite number of conserved charges
introduce selection rules which protect them against decay into lower
energy vortex solutions, and so give them some stability. However the question
of stability of such solutions 
under small perturbations is  a much more complex issue, and
deserves further studies.   We should point out however that the static
Hamiltonian associated to 
  \rf{action} is positive definite for $M^2>0$,
  $e^2<0$, $\beta<0$ and $\beta\,e^2\geq 1$. Therefore, our static
  vortices, saturating the Bogomolny bound (see \rf{hcbound}), have
  the smallest possible  
  energy in such theory, for that range of coupling constants. That
  fact makes us believe
  that for sufficiently large $M^2$, the vortices should be
  stable under small perturbations. 

If one chooses in \rf{currents}, the functional as $G=-i\,4\,\(1+\mid  
u\mid^2\)^{-1}$, and if one takes into 
account \rf{bogo3} one gets that the corresponding conserved current
is  
\be
J_{\mu}=
-i\,4\,M^2\,\frac{\left[u\,\partial_{\mu}u^*-u^*\,\partial_{\mu}u\right]}
{\(1+\mid u\mid^2\)^2} 
\ee
which corresponds to the Noether current for the theory \rf{action}
associated to the symmetry $u\rightarrow e^{i\alpha}\,u$, after the
conditions \rf{constraint} and \rf{bogo3} are imposed. Evaluating the
charge per unit length for the solution \rf{vortexwave} one gets
\be
Q=\int dx^1dx^2 \,J_0 = \varepsilon_2\,8\,\pi\,M^2\,k\,a^2\, I\(\mid n\mid\)
\ee
Therefore, the contribution to the energy per unit length
\rf{energyvortexwave} coming from 
the plane wave is proportional to the $U(1)$ Noether charge it gives to the
solution.

The spectrum of energy we have got is therefore very similar to that of
the so-called Q-lumps constructed in \cite{leese} for the $2+1$
dimensional $CP^1$ model with a potential. See \cite{manton} for more
details, \cite{wojtek} for a related  model, and also \cite{adamqball}
for $Q$-balls solutions on $S^3$. However, the
mechanisms involved here are quite different. We have a four
dimensional  theory, without a potential term depending on $\mid u\mid$
only, and our vortices are not spinning but instead have waves traveling along
them with the speed of light. In addition, we again point out that our
results are true for the four dimensional $CP^1$ model without a
potential, but subjected to the constraint \rf{constraint}. 

The theory \rf{action} is invariant under the internal $SO(3)$
symmetry of rotations of the ${\vec n}$ field. It then possesses three
conserved Noether currents associated to such symmetry. One of them is
that associated to the symmetry $u\rightarrow e^{i\alpha}\,u$, discussed
above. However, when the contraint \rf{constraint} is imposed the
reduced model gets an infinite number of conserved currents given by
\rf{currents}. Such currents can not be of the Noether
type\footnote{Except for those three currents associated to the global
  $SO(3)$ symmetry of \rf{action}, contained in \rf{currents}.}
because the reduced system of equations, namely $\partial^2 u =0$ and
$\(\partial_{\mu}u\)^2=0$, does not possess a Lagrangian and so the
Noether theorem does not apply. Such currents can be understood as
hidden symmetries of the reduced model by constructing them using a
zero curvature representation as explained in section 6.1.1 of
\cite{afs}. They are associated to an infinite dimensional
non semi-simple Lie algebra made of the semi-direct product of $SU(2)$
with abelian ideals transforming under infinite dimensional
representations of $SU(2)$.  It is important to understand the role
such symmetries may have in the 
extension of the Skyrme-Faddeev model given by the theory \rf{action}. 

Another  point concerning the enlargement of symmetries is the
following. Even though the theory 
\rf{action} is not scale invariant, the conditions \rf{constraint} and
\rf{bogo3} lead to a sub-model which  equations of motion
are scale invariant. The condition $\(\partial u\)^2=0$ is conformally
invariant but $\partial^2u=0$ is only scale invariant in four
dimensions.

Finally, we would like to comment on some important points of our
construction which relate to the pure $SU(2)$ Yang-Mills theory. 
Notice that by using the
  Cho-Faddeev-Niemi 
  decomposition \cite{chofn,fn} of the $SU(2)$ gauge field ${\vec
  A}_{\mu}$, and 
  assuming that at low energies 
  all degrees of freedom get frozen except for those associated to the
  ${\vec n}$ field, then 
${\vec A}_{\mu}=\partial_{\mu}{\vec n}\wedge {\vec n}$. Our vortices
are solutions of the constraints 
$\partial_{\mu}u\partial^{\mu}u=0$, and also of $\partial^2u=0$. Those
equations imply that the field ${\vec n}$ satisfies $\partial^2{\vec
  n}+\left(\partial {\vec n}\right)^2\,{\vec n}=0$. Therefore the
connection, evaluated on the vortices, satisfies the gauge fixing condition 
\be
\partial^{\mu}{\vec A}_{\mu}=0
\ee
Another point is that, without the use of any condition
(equations of motion or constraint), and working in a Euclidean
space-time, it follows 
that (see eq. \rf{dn2})
\be
{\vec A}_{\mu}\cdot
{\vec A}^{\mu} ={\cal H}_c/M^2 
\ee
where ${\cal H}_c$ is given in \rf{hcdef}. Therefore as explained in
\rf{hcbound}, the equations \rf{bogo4} then imply that our vortex solutions
minimize locally 
the quantity ${\vec A}_{\mu}\cdot {\vec A}^{\mu}$. In fact, its
integral on the plane $x^1\, x^2$ has the smallest possible value when
evaluated on the vortices. Such quantity is used as a gauge fixing
condition and it is believed to signal non trivial structures in gauge
theories \cite{a2}.

 In
order to have the condition \rf{bogo3} satisfied we need of course the
coupling constants $\beta$ and $e^2$ to have the same sign. That
implies that the second and third terms in \rf{action} have opposite
signs. We are then in a situation which differs from that considered
in \cite{glad} where the signs of the three terms in \rf{action} were
chosen to give positive contributions to the static energy. See also 
\cite{sawado} for a discussion on solutions of that type of model.  We are
however in agreement with the situation in \cite{gies}. Indeed,
consider the action \rf{action} in four dimensional Euclidean space
and take $\beta$ and $e^2$ to be negative. We then have the same sign
structure of the effective action given in eq. (14) of
\cite{gies}. In fact, the condition \rf{bogo3} corresponds to the following
relation among the parameters of \cite{gies}
\be
\ln\frac{k}{\Lambda} =-6\,\pi^2\,\frac{\(4-\alpha\)}{\alpha\,g^2}
\lab{condt}
\ee
where $g^2$ is the gauge coupling constant, $\alpha$ a gauge fixing
parameter, $\Lambda$ and $k$ are the UV and IR 
cutoffs respectively.  The values for $\alpha$ in
\cite{gies} are such that $\(4-\alpha\)/\alpha$ is positive, and so
\rf{condt} is indeed 
consistent with $k<\Lambda$.  However, the relation \rf{condt} has to
be read in 
its proper context, i.e. the physical limit corresponds to
$k\rightarrow 0$, and the perturbative results of \cite{gies} may not
be applicable in that regime. It would be very interesting to
investigate if the condition \rf{bogo3} remains consistent with the
re\-nor\-ma\-li\-za\-tion group flows in the IR limit. That could
clarify if the 
solutions calculated in this paper can play a role in the low
energy limit of the pure $SU(2)$ Yang-Mills theory. \\

\noindent{\bf Acknowledgments:} The author is very grateful to
Prof. N. Sawado (Tokyo Univ. of Science), Prof. K. Toda (Toyama
Prefectural Univ.), Prof. R. Sasaki (Yukawa Inst. for Theor.
Physics) and Prof. O. Alvarez (Univ. of Miami) for many
helpful discussions and the hospitality at their institutions where part of
this work was done. He is also very grateful to Prof. H. Gies (Jena
Univ.),  Prof. J. Sanchez
Guill\'en (Univ. Santiago de Compostela), and Prof. W.J. Zakrzewski
(Univ. of Durham)    for suggestions and a careful
reading of the manuscript.

\end{document}